\documentclass[aps,prd,notitlepage,showpacs,nofootinbib,superscriptaddress]{revtex4-1}

\usepackage{graphicx}
\usepackage[utf8]{inputenc} 

\usepackage{amsmath}
\usepackage{amssymb}
\usepackage{float}
\usepackage{comment}
\usepackage{slashed}
\usepackage[normalem]{ulem}

\usepackage{booktabs}
\usepackage{mathtools,braket,blkarray}

\usepackage{amsmath}
\usepackage{amssymb}

\usepackage[usenames,dvipsnames]{color}
\usepackage[colorinlistoftodos]{todonotes}
\usepackage[colorlinks=true,citecolor=darkred,urlcolor=darkred, pdfborder={0 0 0}]{hyperref}
\usepackage[normalem]{ulem}

\usepackage{multirow}
\definecolor{darkred}{rgb}{0.6,0,0}
\usepackage[colorinlistoftodos]{todonotes}

\definecolor{linkcolor}{rgb}{0,0,0.5}

\newcommand {\ignore}[1]{}

\def\gsim{\raise0.3ex\hbox{$\;>$\kern-0.75em\raise-1.1ex\hbox{$\sim\;$}}}
\def\lsim{\raise0.3ex\hbox{$\;<$\kern-0.75em\raise-1.1ex\hbox{$\sim\;$}}}

\def\SM{$\mathrm{SU(3)_c \otimes SU(2)_L \otimes U(1)_Y}$ }
\newcommand{\sm}{{Standard Model }}

\usepackage{soul}

\definecolor{mightnightblue}{RGB}{25,25,112}

\definecolor{brown}{rgb}{0.59, 0.29, 0.0}

\def\SM{$\mathrm{SU(3)_c \otimes SU(2)_L \otimes U(1)_Y}$ }
\def\21{$\mathrm{SU(2)_L \otimes U(1)_Y}$}

\def\sm{standard model }

\newcommand{\AddrAHEP}{%
  AHEP Group, Institut de F\'{i}sica Corpuscular --
  C.S.I.C./Universitat de Val\`{e}ncia, Parc Cient\'ific de Paterna.\\
  C/ Catedr\'atico Jos\'e Beltr\'an, 2 E-46980 Paterna (Valencia) - SPAIN}

\begin{document}

\title{\boldmath \color{BrickRed} Scotogenic dark matter in an orbifold theory of flavor}

 \author{Francisco J. de Anda}\email{fran@tepaits.mx}
\affiliation{Tepatitl{\'a}n's Institute for Theoretical Studies, C.P. 47600, Jalisco, M{\'e}xico}

 \author{Ignatios Antoniadis}\email{antoniad@lpthe.jussieu.fr}
\affiliation{Laboratoire de Physique Th\'eorique et Hautes Energies (LPTHE),\\ UMR 7589,
Sorbonne Universit\'e et CNRS, \\ 4 place Jussieu, 75252 Paris Cedex 05, France.} 
\affiliation{ Albert Einstein Center, Institute for Theoretical Physics, University of Bern, \\
       Sidlerstrasse 5, 3012 Bern, Switzerland.
} 

 \author{Jos\'{e} W. F. Valle}\email{valle@ific.uv.es}
 \affiliation{\AddrAHEP}

 \author{Carlos A. Vaquera-Araujo}\email{vaquera@fisica.ugto.mx}
 \affiliation{Consejo Nacional de Ciencia y Tecnolog\'ia, Avenida Insurgentes Sur 1582. Colonia Cr\'edito Constructor, Alcald\'ia Benito Ju\'arez, C.P. 03940, Ciudad de M\'exico, M\'exico}
 \affiliation{Departamento de F\'isica, DCI, Campus Le\'on, Universidad de
 Guanajuato, Loma del Bosque 103, Lomas del Campestre C.P. 37150, Le\'on, Guanajuato, M\'exico}

\begin{abstract}
\vspace{0.5cm}

We propose a flavour theory in which the family symmetry results naturally from a six-dimensional orbifold compactification.
``Diracness'' of neutrinos is a consequence of the spacetime dimensionality, and the fact that right-handed neutrinos live in the bulk.
Dark matter is incorporated in a scotogenic way, as a result of an auxiliary $\mathbb{Z}_3$ symmetry, and its stability is associated to the conservation of a ``dark parity'' symmetry.
The model leads naturally to a ``golden'' quark-lepton mass relation. 

\end{abstract}

\maketitle
\noindent

\section{Introduction}

Two major drawbacks of the \sm is the lack of neutrino mass and dark matter.
 Indeed, the discovery of neutrino oscillations implies that neutrinos have mass \cite{McDonald:2016ixn,Kajita:2016cak,deSalas:2017kay} and the need to supplement the Standard Model (SM).
 Likewise, many models adding particle dark matter to the \sm can be envisaged.
 An interesting idea is that dark matter is the mediator of neutrino mass generation, the corresponding models have been dubbed
 scotogenic~\cite{Ma:2006km,Farzan:2012sa,Hirsch:2013ola,Merle:2016scw,Bonilla:2016diq,Reig:2018mdk,Avila:2019hhv,Kang:2019sab,Leite:2019grf,Leite:2020wjl}.
An even tougher challenge in particle physics is understanding flavor, i.e. the pattern of fermion mixings as well as their mass hierarchies.
The latter suggests extending the \sm by the imposition of a family symmetry in order to provide them a non-trivial structure. 
However, there are just too many possibilities to choose from~\cite{Chen:2015jta,Ishimori:2010au}. 

Rather than imposing a flavour symmetry in an \textit{ad hoc} fashion, it could arise from extra dimensions \cite{Antoniadis:1998ig}.
Here we assume the spacetime to be 6-dimensional, where the extra dimensions are orbifolded as $\mathbb{T}^2/\mathbb{Z}_2$ from which, after compactification,
emerges an $A_4$ flavour symmetry~\cite{Babu:2002dz,deAnda:2018oik,deAnda:2018yfp}. 
We assume a similar setup as in \cite{deAnda:2019jxw,deAnda:2020pti}. 
However, there are crucial differences from the previous work, where the neutrinos were Majorana-type, with masses arising from the type-I seesaw mechanism.
The model studied in this paper does not allow Majorana masses for neutrinos, since they are 6-dimensional chiral fermions, therefore they are fixed to be Dirac type when reduced to 4-dimensions.
Due to an auxiliary triality or $\mathbb{Z}_3$ symmetry, the small neutrino masses are generated only at one-loop level.
All the mediators running in the loop are "dark", i.e. charged under a $\mathbb{Z}_2$ symmetry which makes the lightest of them stable and hence a Dark Matter candidate.
Therefore this is a realization of the Dirac scotogenic mass generation mechanism~\cite{Farzan:2012sa,Bonilla:2016diq,Reig:2018mdk,Leite:2019grf,Leite:2020wjl}.
The presence of the family symmetry leads naturally to a ``golden'' quark-lepton mass relation~\cite{Morisi:2011pt,King:2013hj,Morisi:2013eca,Bonilla:2014xla,Bonilla:2017ekt,Reig:2018ocz}.

This letter is organized as follows. In Sec.~\ref{Sec:setup} we briefly describe how we obtain the $A_4$ family symmetry from extra dimensions, in Sec.~\ref{sec:basic-framework}
we discuss the basic framework and quantum numbers. Fermion masses are discussed in Sec.\ref{sec:fermion-masses}, including both the charged fermions as well as the
scotogenic neutrino masses in Sec.\ref{subsec:scot-neutr-mass}. 

\section{$A_4$ family symmetry from extra dimensions}
\label{Sec:setup}

We assume the spacetime manifold as $\mathcal{M}=\mathbb{M}^4\times (\mathbb{T}^2/\mathbb{Z}_2)$, where the torus $\mathbb{T}^2$ is defined by the relations
\begin{equation}\begin{split}
z&=z+1,\\
z&=z+\omega,\\
z&=-z,
\label{eq:orbtra}
\end{split}\end{equation}
where we rescale the original radii of the torus as $2\pi R_1\Rightarrow 1$ and $2\pi R_2\Rightarrow 1$ and adopt the complex coordinate notation $z=x_5+ix_6$. 
The twist of the torus is the cube root of unity  $\omega=e^{i\theta}=e^{2i\pi/3}$.
There are four fixed points under these orbifold transformations, that define four invariant 4-dimensional branes 
\begin{equation}
\bar{z}=\left\{0,\ \frac{1}{2},\ \frac{\omega}{2},\ \frac{1+\omega}{2}\right\}.
\label{eq:orgbro}\end{equation}
After orbifold compactification, a remnant symmetry of the set of branes is inherited from the Poincar\'e invariance of the extra dimensional part of the manifold  \cite{Altarelli:2006kg,Adulpravitchai:2009id,Adulpravitchai:2010na}.
The transformations that permute the four branes leaving the whole brane-set $\bar{z}$ invariant are 
\begin{equation}
S_1:z\to z+1/2,\ \ \ S_2:z+\omega/2,\ \ \ R:z\to\omega^2 z,
\label{eq:remsym}
\end{equation}
which are just translations and rotations of the extra dimensional coordinates. 
Among this set of transformations there are only two independent ones, since $S_2=R^2\cdot S_1\cdot R$. 
These symmetry transformations relate to the $A_4$ generators through the identification $S=S_1,\ T=R,$ satisfying  
\begin{equation}
S^2=T^3=(ST)^3=1,
\end{equation}
which is the presentation for the $A_4$ group. Therefore, the brane set is invariant under $A_4$ transformations, which are a subset of the Extra Dimensional part of the Poincar\`e group. 

The fields located on the branes will transform naturally under the remnant $A_4$ group. Following Refs.~\cite{deAnda:2019jxw,deAnda:2020pti} we identify this remnant symmetry as a family symmetry. 
The orbifold compactification also fixes the possible representations for the fields localized on the branes. One can write the $S,T$ as matrices acting on $(\bar{z}_1,\bar{z}_2, \bar{z}_3,\bar{z}_4)^T=\big(0,1/2,\omega/2,(1+\omega)/2\big)^T$ as 
\begin{equation}
S=\left(\begin{array}{cccc}0&1&0&0
\\ 1&0&0&0
\\ 0 &0&0&1
\\ 0&0&1&0\end{array}\right), \ \ \ 
T=\left(\begin{array}{cccc}1&0&0&0
\\ 0&0&1&0
\\ 0&0&0&1
\\ 0&1&0&0\end{array}\right),
\end{equation}
so that the 4 branes transform as a reducible representation $\textbf{4}$ of $A_4$.
One can decompose the $\textbf{4}$ into irreducible representations $\textbf{4}\to \textbf{3}+\textbf{1}$. This can be made explicit with a basis change of the $S,T$ transformations, through a unitary transformation $V$, namely
\begin{equation}
S\to V^\dagger S V=\left(\begin{array}{cc} S_{3}&0 \\ 0&1\end{array}\right),\ \ \ T\to V^\dagger T V=\left(\begin{array}{cc} T_{3}&0 \\ 0&1\end{array}\right),
\end{equation}
where $S_3,T_3$ are the usual $3\times 3$ matrix representation of $A_4$. In this way the 4 dimensional representation inherited from the branes can be explicitly decomposed as $\textbf{4}\to \textbf{3}+\textbf{1}$.
Therefore, the fields on the branes must transform under the flavour group $A_4$ as the irreducible representations $\mathbf{3}$ or $\mathbf{1}$, depending on the location of each component of the field \cite{Bazzocchi:2009pv,deAnda:2018oik}. A brane field $F_i(x,z)$ transforming as a $\mathbf{3}$ of the remnant symmetry $A_4$ would be written as a sum of 4-d fields located on different branes
\begin{equation}
F_i(x,z)=\sum_{j=1,2,3} \delta^2(z-\bar{z}_j) V^\dagger_{ij} F_i(x),
\label{eq:trip}
\end{equation}
for $i=1,2,3$ (no sum implied on $i$) which are the components of the triplet. Therefore different components of the triplet are located at different branes, and they transform into each other by the $A_4$ transformations just as the branes do.
The singlet $f(x,z)$ 
\begin{equation}
f(x,z)=\sum_{j=1,2,3,4} \delta^2(z-\bar{z}_j) V^\dagger_{4j} f(x)=\sum_{j=1,2,3,4}\frac{1}{2}\delta^2(z-\bar{z}_j) f(x),
\label{eq:sing}
\end{equation}
is equally located on the four branes.

Notice that, a priori, one can start with either a triplet or a singlet, if one locates their components as in Eqs.~(\ref{eq:trip}),~(\ref{eq:sing}), respectively.
  Concerning fields in the bulk, for consistency all 6-dimensional fields should also transform under some irreducible representation of the $A_4$ remnant symmetry.
  This way one can also localize them consistently with $A_4$ irreducible representations~\cite{Burrows:2009pi}.

\section{Basic framework}
\label{sec:basic-framework}

The field content of our model and its transformation properties is specified in Table \ref{tab:fieldspv}.
The model contains the usual \sm fermions $L,d^c,e^c,Q,u^c$, extended by three right handed neutrinos $\nu^c$.
All the scalar fields are localized on the branes and consequently transform as flavour triplets.
The model includes two electroweak scalar doublets $H_u,H_d$, together with an \SM singlet scalar $\sigma$ that obtains a vacuum expectation value (VEV) above the electroweak scale,
breaking the $A_4$ flavour symmetry.
We assume the existence of an independent shaping symmetry $\mathbb{Z}_3\times\mathbb{Z}_2$. The scalars are charged under the $\mathbb{Z}_3$ symmetry, so that $H_d$ only couples to down-type fermions (charged leptons and down quarks), $H_u$ couples only to up-quarks and $\eta$ only couples to neutrinos,
{\it i.e.} the $\mathbb{Z}_3$ symmetry fixes the the Higgs couplings to be selective, as in type II 2HDM.
A very important feature for us is that the $\mathbb{Z}_3$ symmetry forbids a renormalizable coupling that would give tree-level Dirac masses to neutrinos.
On the other hand, the unbroken $\mathbb{Z}_2$ symmetry stabilizes the lightest ``dark'' field.  In order to generate scotogenic neutrino masses, two sets of inert scalars $\eta$ and $\chi$ are included, together with the electroweak singlet fermions $S$.
\begin{table}[h]
\centering
\footnotesize 
\begin{tabular}{ |cc|ccc|cccccccc|}
\hline
\textbf{Field} &$\qquad$ & $SU(3)$ & $SU(2)$ & $U(1)$ &$\qquad$  &$A_4$ &$\qquad$& $\mathbb{Z}_3$&  $\qquad$& $\mathbb{Z}_2$&$\qquad$& Localization \\
\hline
$L$ & & $\mathbf{1}$ & $\mathbf{2}$ & $-1/2$ & & $\mathbf{3}$  & & $\omega^2$&&$1$ & &Brane\\
$d^c$ && $\bar{\mathbf{3}}$ & $\mathbf{1}$ & $1/3$ & & $\mathbf{3}$  & & $\omega$&&$1$&& Brane\\
$e^c$ &&$\mathbf{1}$ & $\mathbf{1}$ & $1$ & & $\mathbf{3}$  & & $\omega$&&$1$&& Brane\\
$Q$ && $\mathbf{3}$ & $\mathbf{2}$ & $1/6$ & & $\mathbf{3}$  & & $\omega^2$&&$1$&& Brane\\
$u^c$ && $\bar{\mathbf{3}}$ & $\mathbf{1}$ & $-2/3$ & & $\mathbf{3}$ & & $\omega^2$&&$1$&& Brane \\
$\nu^c_i$ && $\mathbf{1}$ & $\mathbf{1}$ & $0$ & & $\mathbf{1}$   & & $1$&&$1$&&Bulk \\
$T^c$ && $\bar{\mathbf{3}}$ & $\mathbf{1}$ & $-2/3$ & & $\mathbf{1}$ & & $\omega$&&$1$&& Brane \\
$T$ && $\mathbf{3}$ & $\mathbf{1}$ & $2/3$ & & $\mathbf{1}$ & & $\omega^2$&&$1$&& Brane \\
\hline
$H_u$ && $\mathbf{1}$ & $\mathbf{2}$ & $1/2$ & & $\mathbf{3}$ & & $\omega^2$&&$1$&& Brane \\
$H_d$  && $\mathbf{1}$ & $\mathbf{2}$ & $-1/2$ & & $\mathbf{3}$  & & $1$&&$1$&&Brane\\
$\sigma$ && $\mathbf{1}$ & $\mathbf{1}$ & $0$ & & $\mathbf{3}$ & & $\omega$&&$1$&& Brane\\
\hline
$S$ && $\mathbf{1}$ & $\mathbf{1}$ & $0$ & & $\mathbf{3}$   & & $\omega$&&$-1$&&Brane\\
\hline
$\eta$ && $\mathbf{1}$ & $\mathbf{2}$ & $-1/2$ & & $\mathbf{3}$ & & $1$&&$-1$&& Brane \\
$\chi$  && $\mathbf{1}$ & $\mathbf{1}$ & $0$ & & $\mathbf{3}$  & & $\omega^2$&&$-1$&&Brane\\
\hline
\end{tabular}
\caption{Field content of the model.}
\label{tab:fieldspv}
\end{table}

Notice that, except for the $\nu^c$, all fields are assumed to live in the branes. Therefore, the $L,d^c,e^c,Q,u^c,S$ fields are 4-d Weyl left-fermions and $H_u,H_d,\eta,\chi$ are 4-d scalars.
In the 4d branes, we have the SM fermions plus a vector-like quark, therefore gauge anomalies are canceled just as in the standard model.
  The only 6d fermion is the right handed neutrino which is a gauge singlet, therefore it does not contribute to any anomaly.
These fields on the branes behave as standard 4-d fields, and we restrict them to have only renormalizable couplings in the low energy theory.

In contrast, the $\nu^c$ are located in the bulk and are assumed to be 6-d Weyl fermions, each of which decomposes into a left-right 4-d Weyl fermion pair.
The trivial boundary conditions on the orbifold do not allow a zero mode for the right-handed part.
Therefore, after compactification, each of them decomposes into a massless 4-d Weyl left fermion, plus the corresponding Kaluza-Klein (KK) tower.
 If we were to assume this model to be valid up to the Planck scale, it should also be free of gravitational anomalies.
  The fermion fields on the brane are the \sm fermions plus vector-like quarks, therefore gravitational anomaly cancellation happens just as in the standard model.
  The only bulk fields are the three 6-d chiral fermions $\nu^c_i$ which have three 4-d chiral fermions as zero modes.
  These would generate a gravitational anomaly.
  To cancel it one can easily add three 6-d chiral fermions $\bar{\nu}^c_i$ with opposite chirality from $\nu^c_i$. 
  Furthermore, we assume it to be nontrivially charged under both discrete symmetries $\mathbb{Z}_3$ and $\mathbb{Z}_2$, so as not to couple directly to the SM fields, hence preserving the phenomenology discussed below.

\section{Fermion masses}
\label{sec:fermion-masses}

\subsection{Quarks and charged leptons}
\label{subsec:QL}

The fermions in the bulk are 6-d Weyl fermions whose dimensionality is $[u,\nu]=5/2$.
They can have an explicit mass term which does not affect the zero modes at low energies. Their couplings to the brane fields come from the 6-d lagrangian
\begin{equation}
\mathcal{L}_6=\delta^2(z-\bar{z}) {\frac{\tilde{y}_{i}^{\nu_2}}{\Lambda} S \chi \nu^c_{i}  } ,
\end{equation}
which is suppressed by an effective scale $\Lambda$. The effective 4-d couplings at low energies absorb this scale as $y=\tilde{y} /R\Lambda$, where $R$ denotes the radii of the torus and defines the compactification scale. \\[-.2cm]

The quarks and charged leptons behave in a very similar way as in the model described in~\cite{deAnda:2019jxw,deAnda:2020pti}.
We assume that all dimensionless couplings are real, and that the Higgs fields $H_{u,d}$ and the $\sigma$ get a VEV that break CP spontaneously. We parametrize them as as
\begin{equation}\begin{split}
\braket{H_u}=v_u\left(\begin{array}{c}\epsilon_1^u e^{i\phi_1^u}\\ \epsilon_2^u e^{i\phi_2^u}\\ 1\end{array}\right),\ \ \ \braket{H_d}=\frac{v_u}{\tan\beta} e^{i\phi^d}\left(\begin{array}{c}\epsilon_1^d e^{i\phi_1^d}\\ \epsilon_2^d e^{i\phi_2^d}\\ 1\end{array}\right),\ \ \ \braket{\sigma}=v_u r e^{i\phi^\sigma}\left(\begin{array}{c}\epsilon_1^\sigma e^{i\phi_1^\sigma}\\ \epsilon_2^\sigma e^{i\phi_2^\sigma}\\ 1\end{array}\right),
\end{split}\end{equation}
where $r=\braket{|\sigma|}/v_u$ defines the scale of the $A_4$ breaking and can be very large.
After compactification the Yukawa couplings for the up quarks are assumed to be
\begin{equation}
\begin{split}
\mathcal{L}_q = 
y_{1}^u (Q H_u u^c)_{1}+y_{2}^u (Q H_u u^c)_{2}+y^T Q H_d^\dagger T^c+\tilde{y}^T\sigma^\dagger u^c T+M_T TT^c  +\text{h.c.},
\label{eq:yukq}
\end{split}
\end{equation}
where the $()_{1,2}$ denote the two possible $\textbf{3}\times \textbf{3}\to \textbf{3}_{1,2}$ contractions.
Notice the presence of one pair of exotic vector-like quarks.
The reason for adding them is that
the up quark mass matrix coming from the first two terms does not, by itself, have enough freedom to generate a realistic CKM matrix.
  The role of the extra vector-like up-type quarks is to generate a viable mixing.
  The added free parameters ($y^T,\tilde{y}^T,M_T$) are enough to account for the full structure of the CKM matrix, as shown in Ref.~\cite{Morisi:2013eca}.
In fact, the presence of vector-like fermions has also been recently suggested in Ref.~\cite{Crivellin:2020ebi}.
The explicit form of the up-type quark mass matrices is given in the Left-Right convention as
\begin{equation}\begin{split}
M_u&=v_u\left(\begin{array}{cccc} 0 & y_1^u\epsilon_1^u e^{i\phi_1^u}& y_2^u \epsilon_2^u e^{i\phi^u_2 }& y^T \cot\beta \epsilon^d_1 e^{i\phi^d_1}\\
 y_2^u \epsilon_1^u e^{i\phi_1^u}& 0 &  y_1^u  & y^T \cot\beta \epsilon^d_2 e^{i\phi^d_2}\\
 y_1^u \epsilon_2^u e^{i\phi^u_2}& y_2^u  &0& y^T \cot\beta\\
 \tilde{y}^Tr\epsilon^\sigma_1e^{-i\phi^\sigma_1} &  \tilde{y}^Tr\epsilon^\sigma_2e^{-i\phi^\sigma_2} & \tilde{y}^Tr & M_T e^{i(\phi^\sigma-\phi^d)}/v_u\end{array}\right),
\label{eq:umassmat}
\end{split}\end{equation}
where the phase $e^{i(\phi^\sigma-\phi^d)}$ can be reabsorbed and the ratio $M_T/v_u$ is assumed to be very large $M_T/v_u\gg 1$.

The Yukawa couplings for the charged leptons and down-type quarks, after compactification become
\begin{equation}
\begin{split}
\mathcal{L}_l=y_1^d(Q d^c H_d)_1+y_2^d(Q d^c H_d)_2+y_1^e(L e^c H_d)_1+y_2^e(L e^c H_d)_2+\text{h.c.}
\label{eq:yukg}
\end{split}
\end{equation}
These terms generate the mass matrices
\begin{equation}\begin{split}
M_d=v_u\cot\beta\left(\begin{array}{ccc} 0 & y_1^d\epsilon_1^d e^{i(\phi_1^d-\phi_2^d)}& y_2^d \epsilon_2^d \\
 y_2^d\epsilon_1^d e^{i(\phi_1^d-\phi_2^d)}& 0 &  y_1^d\\
 y_1^d\epsilon_2^d & y_2^d&0\end{array}\right),\ \ \ 
M_e=v_u\cot\beta\left(\begin{array}{ccc} 0 & y_1^e\epsilon_1^d e^{-i(\phi_1^d-\phi_2^d)}& y_2^e \epsilon_2^d \\
 y_2^e\epsilon_1^d e^{-i(\phi_1^d-\phi_2^d)}& 0 &  y_1^e\\
 y_1^e\epsilon_2^d & y_2^e&0\end{array}\right).
\label{eq:dmassmat}
\end{split}\end{equation}
Since the above interactions leading to down-type-quark and lepton masses involve the same Higgs doublet, all of which are $A_4$ triplets, the golden relation between charged lepton and down quark
masses~\cite{Morisi:2011pt,King:2013hj,Morisi:2013eca,Bonilla:2014xla,Bonilla:2017ekt,Reig:2018ocz} emerges as a prediction from the family symmetry
\begin{equation}
\frac{m_\tau}{\sqrt{m_\mu m_e}}\approx \frac{m_b}{\sqrt{m_s m_d}},
\label{eq:golden}
\end{equation}

 This is a prediction for the pattern of fermion masses, relating down-quark and charged lepton masses, despite the lack of an underlying unification gauge group~\cite{Morisi:2011pt,King:2013hj,Morisi:2013eca,Bonilla:2014xla,Bonilla:2017ekt,Reig:2018ocz}.
 This relation is stable under renormalization group evolution~\cite{Antusch:2013jca}, as it involves only ratios of fermion masses. 
 It has been explicitly shown to provide an excellent description of the current experimental data at the $M_Z$ scale \cite{deAnda:2019jxw,deAnda:2020pti}. \\[-.2cm]

The first two terms of each of Eqs.~(\ref{eq:yukq}),(\ref{eq:yukg}), involving only the standard quark families, can fit the quark sector within $10\%$ accuracy.
  The addition of the vector-like quarks brings in free parameters that help fit the quark flavour observables.
  All in all, the quark and charged lepton sector has 16 arbitrary real dimensionless parameters $(y_{1,2}^{u,d,e},y^T,\tilde{y}^T,\epsilon_{1,2}^{u,d,\sigma},\tan\beta,r )$ plus
  extra phases describing the alignment of the Higgs bosons. 
 In Ref.~\cite{Morisi:2013eca} it has been shown explicitly that the above structure of quark mass matrices provides an excellent fit to all the quark flavour observables.

\subsection{Scotogenic neutrino masses}
\label{subsec:scot-neutr-mass}

Besides the \sm fermions and the enlarged scalar states described above, the particle content is minimally extended by a dark sector consisting of a fermion $S$ and two
scalars\footnote{In fact the scalar $\eta$ plays the role of the $H_\nu$ scalar present in~\cite{deAnda:2019jxw,deAnda:2020pti}, but now belongs to the ``odd'' dark matter sector.}
$\eta,\chi$.
In the low energy theory, the Yukawa couplings for neutrinos and fermion electroweak singlets are
\begin{equation}
\begin{split}
\mathcal{L}_\nu&= y_{1,2}^{\nu_1}(L \eta^\dagger S )_{1,2} +y_1^{\nu_2}(S \chi)_{1}\nu_1^c+y_2^{\nu_2}(S \chi)_{1}\nu_{2}^c+y_3^{\nu_2}(S \chi)_{1}\nu_{3}^c+y^{\nu_3} \sigma SS+\text{h.c.}
\label{eq:ferdars}
\end{split}
\end{equation}
Among the terms in the scalar potential, after compactification we can find the following trilinear couplings:
\begin{equation}
\begin{split}
\mathcal{L}_t/\mu&= y_{1,2}^{t_1} (H_u \eta\chi^\dagger )_{1,2}\  +y^{t_2}\sigma^3+y^{t_3}_{1,2}(\sigma H_u H_d )_{1,2}+y^{t_4}\chi^2\sigma^\dagger+\text{h.c.}
\label{eq:tril}
\end{split}
\end{equation}
{These terms are crucial to ensure the mixing of the neutral components of $\eta$ and $\chi$ and to break any possible degeneracy of the scalar mass eigenstates mediating the neutrino mass generation mechanism.}

\begin{figure}[!h]
\centering
\includegraphics[width=0.6\textwidth]{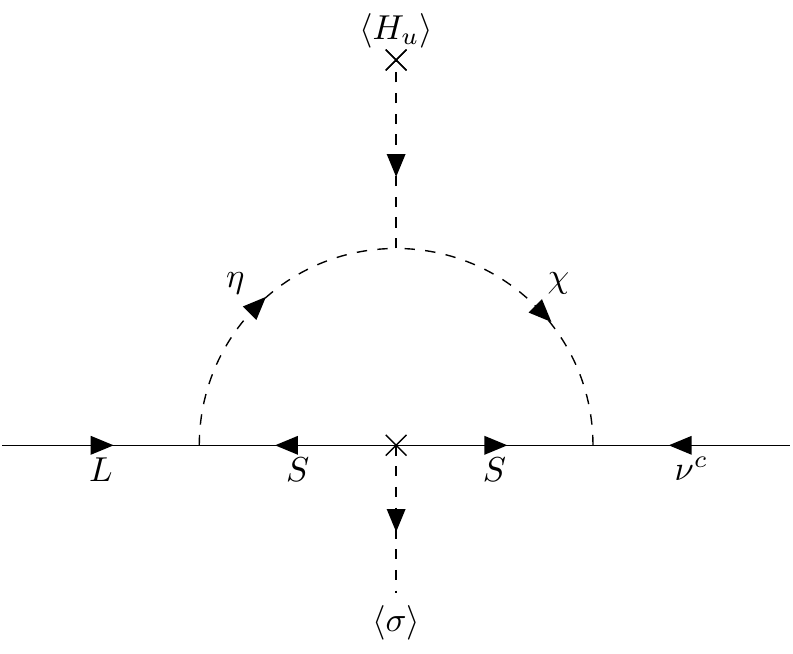}
\caption{Neutrino Dirac masses from ``scotogenic'' loop diagram.}
\label{fig:dirac-scoto}
 \end{figure}

Note that, by choosing the ``right-handed'' neutrinos to be 6-d fermions, even if they are Majorana, their 4-dimensional zero modes are Dirac-type. Moreover, thanks to the auxiliary symmetries in our model, neutrinos are massless at the tree level.
Indeed, the couplings in Eq.~(\ref{eq:ferdars}), together with the first trilinear scalar coupling in Eq.~(\ref{eq:tril}), generate Dirac neutrino masses at one loop from the diagram in Fig.~\ref{fig:dirac-scoto} in a ``scotogenic'' fashion \cite{Ma:2006km,Farzan:2012sa,Hirsch:2013ola,Merle:2016scw,Bonilla:2016diq,Reig:2018mdk,Avila:2019hhv,Kang:2019sab,Leite:2019grf,Leite:2020wjl}.

After spontaneous symmetry breaking, the electrically neutral components of the $A_4$ triplets $\eta$, denoted $\eta^0$ and the electroweak singlets $\chi$ mix into a total of six complex neutral scalars. Assuming trivial $CP$ \footnote{Here we understand trivial $CP$ as the generalized $CP$ transformations of the fields that render all couplings real. See \cite{deAnda:2019jxw,deAnda:2020pti} for details.}, the mass matrix in the basis $(\chi_\gamma,\eta^0_\delta)$, with $\gamma,\delta=1,2,3$ is real and symmetric and can be diagonalized by an orthogonal transformation that can be written in blocks as
\begin{equation}
\left(\begin{array}{c}
\phi_{A\alpha} \\ \phi_{B\alpha}
\end{array}\right)=\left(\begin{array}{cc}
U^{A\chi}_{\alpha\gamma}&U^{A\eta}_{\alpha\delta} \\ U^{B\chi}_{\alpha\gamma}& U^{B\eta}_{\alpha\delta}
\end{array}\right)\left(\begin{array}{c}
\chi_{\gamma} \\ \eta^0_{\delta}
\end{array}\right),
\end{equation}
with $\alpha=1,2,3$. By orthogonality, the $3\times3$ blocks in the above matrix are subject to the relations
\begin{equation}
\begin{split}
&(U^{A\chi})^T U^{A\chi}+(U^{B\chi})^T U^{B\chi}=1,\\
&(U^{A\eta})^T U^{A\eta}+(U^{B\eta})^T U^{B\eta}=1,\\
& (U^{A\chi})^TU^{A\eta}+(U^{B\chi})^TU^{B\eta}=0.\\
\end{split}
\end{equation}

 Neutrinos acquire a small Dirac mass generated through the scotogenic loop.
 To first approximation one gets
\begin{equation}
  \label{eq:scotomass}
m^{\nu}_{ij}=\frac{1}{16\pi^2}\sum_{k\alpha\gamma\delta}(y^{\nu_1}_{1,2})^{\delta}_{ik}M_{k}(y^{\nu_2}_{1,2,3})^{\gamma}_{k  j}\left[(U^{A\chi})^T_{\gamma\alpha}U^{A\eta}_{\alpha\delta}\frac{m_{A\alpha}^2}{m_{A\alpha}^2-M^2_k}\ln \frac{m_{A\alpha}^2}{M^2_k}+(U^{B\chi})^T_{\gamma\alpha}U^{B\eta}_{\alpha\delta}\frac{m_{B\alpha}^2}{m_{B\alpha}^2-M^2_k}\ln \frac{m_{B\alpha}^2}{M^2_k}\right],
\end{equation}
where $M_k$ with $k=1,2,3$ are the eigenvalues of the mass matrix  for the Fermion Singlets $S$,
 $m_A$ ($m_B$) are the physical masses of the $\phi_A$ ($\phi_B$) complex neutral scalars,  and the effective Yukawa couplings $y^{\nu_1}_{1,2}$ and $y^{\nu_2}_{1,2,3}$ are given by
\begin{gather}
(y^{\nu_1}_{1,2})^{1}=\begin{pmatrix}
0 & y^{\nu_1}_1 &0 \\
y^{\nu_1}_2  & 0 & 0\\
0& 0  &0
\end{pmatrix}\,,\qquad
(y^{\nu_1}_{1,2})^{2}=\begin{pmatrix}
0 & 0 &y^{\nu_1}_2 \\
0 & 0 & 0\\
y^{\nu_1}_1& 0  &0
\end{pmatrix}\,,\qquad
(y^{\nu_1}_{1,2})^{3}=\begin{pmatrix}
0 & 0 &0 \\
0 & 0 & y^{\nu_1}_1\\
0& y^{\nu_1}_2  &0
\end{pmatrix}\,,\\
(y^{\nu_2}_{1,2,3})^{1}=\begin{pmatrix}
y^{\nu_2}_1 & y^{\nu_2}_2 &y^{\nu_2}_3 \\
0  & 0 & 0\\
0& 0  &0
\end{pmatrix}\,,\qquad
(y^{\nu_2}_{1,2,3})^{2}=\begin{pmatrix}
0  & 0 & 0\\
y^{\nu_2}_1  & y^{\nu_2}_2 &y^{\nu_2}_3 \\
0& 0  &0
\end{pmatrix}\,,\qquad
(y^{\nu_2}_{1,2,3})^{3}=\begin{pmatrix}
0  & 0 & 0\\
0& 0  &0\\
y^{\nu_2}_1  & y^{\nu_2}_2 &y^{\nu_2}_3 
\end{pmatrix}\,.
\end{gather}

Assuming trivial $CP$ symmetry from the start with real dimensionless couplings, and a general complex VEV alignment for $H_u,H_d,\sigma$ which break CP spontaneously together with the flavour symmetry,
the model can describe realistic fermion masses.
Note that the usual ``left'' neutrinos do not couple directly to the ``right-handed'' neutrinos to form a Dirac mass. Instead, this mass is induced radiatively, through Fig.~\ref{fig:dirac-scoto}. 
  Although the structure of the neutrino Yukawa couplings to the dark sector does have texture zeros implied by the family symmetry, these zeros are lost when composing the scotogenic loop.
  As a result the neutrino mass matrix is the most general one, thus eliminating the predictivity of the model in the neutrino sector.

Besides, our model harbors a dark matter candidate, namely the lightest of the fields running in the scotogenic loop.
  These are charged under the $\mathbb{Z}_2$ symmetry, that remains unbroken after spontaneous symmetry breaking.
As a result, the lightest of the complex neutral scalars $\phi_A$, $\phi_B$ or the Majorana electroweak singlet fermions $S$ is stable and a potential dark matter particle.
A discussion on the viability of this kind of dark matter candidate can be found in \cite{Leite:2020wjl}.

 \section{Summary and outlook}
 \label{sec:summary-outlook}

 We have proposed a scotogenic flavour theory in which the $A_4$ family symmetry arises naturally from a six-dimensional spacetime after orbifold compactification.
 For spacetime dimensionality reasons, neutrinos must be Dirac fermions after compactification, since ``right-handed'' components live in the bulk.
This implies the absence of neutrinoless double beta decay, and makes Majorana phases unphysical~\cite{Schechter:1980gr,Schechter:1980gk}.

 Thanks to the imposition of auxiliary ``dark-parity'' and triality symmetries 
 the theory incorporates stable dark matter in a scotogenic way.
 Neutrinos are massless at tree level, with mass calculable from the scotogenic loop in Fig.~\ref{fig:dirac-scoto}.
 While the theory is not predictive enough to shed light on the structure of the quark CKM mixing matrix or the lepton mixing matrix, it does predict in a rather natural way the ``golden'' quark-lepton
 unification formula, Eq.~(\ref{eq:golden}), despite the lack of a unification group~\cite{Morisi:2011pt,King:2013hj,Morisi:2013eca,Bonilla:2014xla,Bonilla:2017ekt,Reig:2018ocz}.
This structural relation of our model constitutes its only genuine flavor prediction.
 
 The present model therefore provides a scotogenic dark matter completion of the orbifold scenario proposed in Ref.~\cite{deAnda:2019jxw,deAnda:2020pti}, retaining its most remarkable prediction,
 namely the ``golden'' quark-lepton mass relation.

  \section*{Acknowledgments}

  This work is supported by the Spanish grants FPA2017-85216-P (AEI/FEDER, UE), PROMETEO/2018/165 (Generalitat Valenciana) and the Spanish Red Consolider MultiDark FPA2017-90566-REDC. I.A. acknowlegdes funding support by a CNRS-PICS grant 07964. CAV-A is supported by the Mexican Catedras CONACYT project 749 and SNI 58928.

\appendix

\section{6-d fermions}

The Clifford algebra needs an $8\times 8$ matrix representation that satisfy
 \begin{equation}
 \{\Gamma^M,\Gamma^N\}=2\eta^{\mu\nu}\mathbb{I}_8,
 \end{equation}
 where one can use the 6-d chiral representation (in terms of the Pauli matrices
 \begin{equation}
 \begin{split}
 \Gamma^0=\left(\begin{array}{cccc}
 0 & \mathbb{I}_2 &0 &0\\ \mathbb{I}_2&0 &0 &0\\ 0&0&0 & \mathbb{I}_2 \\ 0&0&  \mathbb{I}_2  &0
 \end{array}\right),\  \ \ \ \Gamma^i=\left(\begin{array}{cccc}
 0 & \sigma^i &0 &0\\ -\sigma^i&0 &0 &0\\ 0&0&0 & \sigma^i \\ 0&0&  -\sigma^i  &0
 \end{array}\right),\\
  \Gamma^5=\left(\begin{array}{cccc}
 0 & 0 &i\mathbb{I}_2 &0\\ 0&0 &0 &-i\mathbb{I}_2\\ i\mathbb{I}_2&0&0 & 0\\ 0&-i \mathbb{I}_2&  0 &0
 \end{array}\right),\ \ \ \ 
   \Gamma^6=\left(\begin{array}{cccc}
 0 & 0 &\mathbb{I}_2 &0\\ 0&0 &0 &-\mathbb{I}_2\\ -\mathbb{I}_2&0&0 & 0\\ 0&\mathbb{I}_2&  0 &0
 \end{array}\right).
 \end{split}
 \end{equation}
 Furthermore the chiral matrices are useful
 \begin{equation}
 \begin{split}
 \Gamma^{4C}&=i\Gamma^0\Gamma^1\Gamma^2\Gamma^3=\left(\begin{array}{cccc}
 \mathbb{I}_2 & 0 &0 &0\\ 0&-\mathbb{I}_2 &0 &0\\ 0&0&\mathbb{I}_2 & 0 \\ 0&0& 0  &-\mathbb{I}_2
 \end{array}\right),\\
  \Gamma^{6C}&=i\Gamma^5\Gamma^6=\left(\begin{array}{cccc}
 \mathbb{I}_2 & 0 &0 &0\\ 0&\mathbb{I}_2 &0 &0\\ 0&0&-\mathbb{I}_2 & 0 \\ 0&0& 0  &-\mathbb{I}_2
 \end{array}\right),
 \end{split}
 \end{equation}
 which define the 6-d and 4-d chirality correspondingly.
 
 In 6-d the Dirac fermion has 8 components.  We can write them in terms of 4-d Weyl fermions
 \begin{equation}
 \Psi=\left(\begin{array}{c}\Psi^+_R\\ \Psi^+_L \\ \Psi^-_R \\ \Psi^-_L
 \end{array}\right).
 \end{equation}
 Where each  $\Psi^{\pm}$ (eigenstates of $\Gamma^{6C}$) is composed of a left and right 4-d Weyl fermion  (eigenstates of $\Gamma^{4C}$).
 
 The boundary conditions applied on 6-d fermions are
 \begin{equation}
 \begin{split}
 P\mathcal{P}\Psi(x,z)=P\Gamma^{6C}\Psi(x,-z)&\to P\Psi^+_R(x,-z)\\
 &\to P\Psi^+_L(x,-z)\\
 &\to -P\Psi^-_R(x,-z)\\
 &\to -P\Psi^-_L(x,-z).
 \end{split}
 \end{equation}
 
  The 6-d fermion irreducible representations are 6-d Weyl fermions, eigenstates of 
\begin{equation}
\Gamma^7=-\Gamma^0\Gamma^1\Gamma^2\Gamma^3\Gamma^4\Gamma^5\Gamma^6=\Gamma^{4C}\Gamma^{6C}=\left(\begin{array}{cccc}
 \mathbb{I}_2 & 0 &0 &0\\ 0&-\mathbb{I}_2 &0 &0\\ 0&0&-\mathbb{I}_2 & 0 \\ 0&0& 0  &\mathbb{I}_2
 \end{array}\right),
\end{equation}
so that the irreducible irrepresentations are
\begin{equation}
\Psi_{6R}=\left(\begin{array}{c}\Psi^+_R\\ 0 \\ 0 \\ \Psi^-_L
 \end{array}\right)\ \ \ \ \ {\rm or} \ \ \ \ \ \Psi_{6L}\left(\begin{array}{c}0\\ \Psi^+_L \\ \Psi^-_R \\ 0
 \end{array}\right).
\end{equation}

The 6-d fermions contain both left and right parts. This way, just as in the 5-d case, even if one writes a 6-d Majorana mass term, it decomposes into a 4-d Dirac mass term.

\bibliographystyle{utphys}
\bibliography{bibliography}
\end{document}